\documentclass[final,5p,times,twocolumn]{elsarticle}

\usepackage{amssymb}
\usepackage{amsmath}
\usepackage{amsfonts}
\usepackage{color}
\usepackage{epstopdf}
\usepackage{graphicx}
\usepackage{dcolumn}
\usepackage{amsmath}
\usepackage{amssymb}
\usepackage{bm}
\usepackage{natbib}
\usepackage{xcolor}
\usepackage{mathtools}

\newcommand{\K}{\mathcal{K}}

\newcommand{\highlight}{\color{black}}  
\definecolor{darkGreen}{RGB}{0,150,0}

\usepackage{xparse}

\NewDocumentCommand{\INTERVALINNARDS}{ m m }{
	#1 {,} #2
}
\NewDocumentCommand{\interval}{ s m >{\SplitArgument{1}{,}}m m o }{
	\IfBooleanTF{#1}{
		\left#2 \INTERVALINNARDS #3 \right#4
	}{
	\IfValueTF{#5}{
		#5{#2} \INTERVALINNARDS #3 #5{#4}
	}{
	#2 \INTERVALINNARDS #3 #4
}
}
}

\journal{Journal of Neuroscience Methods}

\begin{document}

\begin{frontmatter}

		\author{Eero Satuvuori\corref{cor1}\fnref{label1,label2,label3}}
		\ead{eero.satuvuori@unifi.it}
		\author{Mario Mulansky\corref{cor1}\fnref{label1}}
		\ead{mario.mulansky@isc.cnr.it}
		\author{Andreas Daffertshofer\corref{cor1}\fnref{label3}}
		\ead{a.daffertshofer@vu.nl}
		\author{Thomas Kreuz\corref{cor1}\fnref{label1}}
		\ead{thomas.kreuz@cnr.it}		
		
		\fntext[label1]{Institute for Complex Systems, CNR, Sesto Fiorentino, Italy}
		\fntext[label2]{Department of Physics and Astronomy, University of Florence, Sesto Fiorentino, Italy}
		\fntext[label3]{Amsterdam Movement Sciences (AMS) \& Institute for Brain and Behaviour Amsterdam (iBBA), Faculty of Behavioural and Movement Sciences, Department of Human Movement Sciences, Vrije Universiteit Amsterdam, The Netherlands}

\title{Using spike train distances to identify the most discriminative neuronal subpopulation}

\begin{abstract}

\noindent \textit{Background:}
Spike trains of multiple neurons can be analyzed following the summed population (SP) or the labeled line (LL) hypothesis. Responses to external stimuli are generated by a neuronal population as a whole or the individual neurons have encoding capacities of their own. The SPIKE-distance estimated either for a single, pooled spike train over a population or for each neuron separately can serve to quantify these responses.

\noindent \textit{New Method:} 
For the SP case we compare three algorithms that search for the most discriminative subpopulation over all stimulus pairs. For the LL case we introduce a new algorithm that combines neurons that individually separate different pairs of stimuli best.

\noindent \textit{Results:}
The best approach for SP is a brute force search over all possible subpopulations. However, it is only feasible for small populations. For more realistic settings, simulated annealing clearly outperforms gradient algorithms with only a limited increase in computational load. Our novel LL approach can handle very involved coding scenarios despite its computational ease.

\noindent \textit{Comparison with Existing Methods:}
Spike train distances have been extended to the analysis of neural populations interpolating between SP and LL coding. This includes parametrizing the importance of distinguishing spikes being fired in different neurons. Yet, these approaches only consider the population as a whole. The explicit focus on subpopulations render our algorithms complimentary.

\noindent \textit{Conclusions:}
The spectrum of encoding possibilities in neural populations is broad. The SP and LL cases are two extremes for which our algorithms provide correct identification results.

\end{abstract}


\end{frontmatter}

\section{\label{s:Introduction} Introduction}

The nervous system is believed to employ large populations of neurons to code and broadcast information. Population coding can be considered less vulnerable and, hence, a more reliable and robust manner than coding via single neurons \citep{Berkowitz09}. In neuronal recordings population coding can appear in two ways. First, all the neurons in the recorded population contribute equally \citep{Rolls97}. Patterns of activity within the population are irrelevant for coding as all that matters is whether or not any of the neurons fires. There, the information being conveyed is that of a single spike train generated by the population as a whole. In contrast to this so-called {\em{summed population} (SP) hypothesis}, each neuron may have a unique and distinguishable role \citep{Huber08, QuianQuiroga09}. In this case, the population is best decoded neuron-by-neuron, which is referred to as the {\em{labeled line} (LL) hypothesis} \citep{Houghton11}. \highlight{Examples for the relevance of each coding scheme in experimental data can be found, e.g., in \citeauthor{Panzeri03}, \citeyear{Panzeri03} (SP) and \citeauthor{Reich01a}, \citeyear{Reich01a} (LL).}

When recording a neuronal population after stimulus presentation, usually only some of the neurons encode the stimulus while others might be involved in different tasks or may exhibit a seemingly erratic activity independent of the stimulus. The responses of these non-coding neurons do not contribute to stimulus discrimination but rather act as a noisy disturbance if included in the analysis. We evaluated different methods to distinguish coding from non-coding neurons under either the SP- or the LL-hypothesis. As will be shown below, the two presumptions require different ways for evaluating stimulus discrimination. 

Spike train distances are a useful means to assess neuronal coding by clustering responses to repeated presentations of a given set of stimuli. If the distance is chosen to be sensitive to the distinguishing features in the spike trains, a small distance between responses to the same stimulus and a large distance between responses to different stimuli can be obtained. While this kind of analysis has been mainly carried out for individual neurons (see e.g. \citeauthor{Chichilnisky05}, \citeyear{Chichilnisky05}; \citeauthor{wang07}, \citeyear{wang07}; \citeauthor{tang14}, \citeyear{tang14}), current technical advances (e.g., \citeauthor{Spira13}, \citeyear{Spira13}; \citeauthor{Lewis15}, \citeyear{Lewis15}) allow for studying neuronal coding in simultaneously recorded populations of neurons \citep{Berenyi13, Packer15}. It has been shown that sensory information is typically not localized in individual neurons \citep{Safaai13} but appears to be distributed over larger neuronal populations \citep{Graf11, Arandia17}. However, the coding via individual neurons and the summation of an entire population are the extreme case in a broad spectrum of possibilities \highlight{\citep{Aronov03, Houghton08}}. In fact, recent evidence points at some intermediate scenario in which a comparably small number encodes information not only in a robust but also very efficient way \citep{Olshausen04, Kwan12}. Ince and colleagues (\citeyear{Ince13}) reported that sensory cortical circuits may process information using small but highly informative ensembles consisting of a few privileged neurons. In the context of brain computer interfaces (BCIs), it was found that a reduced set of carefully selected important neurons exceeded BCI performance levels of the full ensemble \citep{Sanchez04}.

The search for an optimal coding population requires fine-tuned analyses under both the SP- and the LL-hypothesis. For these two cases we show how to separate relevant from irrelevant subpopulations by identifying the subpopulation of neurons amongst all possible ones that discriminates best a given set of stimuli.






\section{\label{s:SPIKE}Spike train distances for neuronal decoding}

Spike train distances can measure the extent to which in a coding population repeated presentations of the same stimulus yield similar spike train responses, while different stimuli result in dissimilar responses. To simulate this, we considered the following setup. $N$ neurons are simultaneously recorded upon repeated presentations of different stimuli -- in a real experiment this is typically done with a multi-electrode array. The number of stimuli $S$ and the number of repetitions $R$ yield an overall number of trials by means of $T = S\!\cdot\!R$. Different spike trains are here denoted as $t_{n,s,r}$ with $n=1,\dots,N$, $s = 1,\dots,S$ and $r = 1,\dots,R$ indexing neurons, stimuli, and repetitions, respectively. Across simulations we selected a subset of neurons to be the coding subpopulation. The goal was, hence, to identify that subset, i.e. the neuronal subpopulation that collectively could distinguish between stimuli.

Spike train distances quantify the similarity of neuronal activity based on rate and timing within spike trains (see e.g. \citeauthor{Houghton11}, \citeyear{Houghton11}; \citeauthor{Kreuz11b}, \citeyear{Kreuz11b}; \citeauthor{Victor15}, \citeyear{Victor15}). Over the years, many different distances have been proposed, including time-scale dependent measures such as the Victor-Purpura distance \citep{Victor96} or the van Rossum distance \citep{VanRossum01} but also time-scale independent approaches like the ISI-distance \citep{Kreuz07c, Kreuz09} and the SPIKE-distance \citep{Kreuz11, Kreuz13}. Here, we employed the SPIKE-distance $D$ \citep{Kreuz13} as it offers the possibility of time-scale and parameter-free assessments via the relative spike timing between spike trains normalized to the local firing rates \citep{Satuvuori18}. The smaller its value, the more similar the spike trains are, with $D = 0$ indicating identical spike trains. A detailed description of the SPIKE-distance can be found in the Appendix. 

Neuronal coding can be assessed by determining the matrix of pairwise spike train distances over all trials (see e.g. \citeauthor{Chichilnisky05}, \citeyear{Chichilnisky05}; \citeauthor{wang07}, \citeyear{wang07}; \citeauthor{tang14}, \citeyear{tang14}). How do these distances cluster in response to different stimuli? Identifying clusters depends on the presumed type of encoding. As said, we distinguish between the SP- and the LL-hypotheses, i.e. we either combine all neurons into a single population or treat each neuron separately. In both cases, we determined distance matrices and estimated their stimulus discrimination performance to quantify how a subpopulation succeeds in discerning different stimuli.

For the summed population case, we compared three fundamentally different algorithms for finding the population that is able to most efficiently discriminate between a set of stimuli. (i) For comparably small neuronal populations one can perform a brute force search in which pairwise distance matrices and their stimulus discrimination performance are calculated for all possible subpopulations.
(ii) A gradient algorithm used by \citet{Ince13} relies on a restricted number of subpopulations: from a given starting subpopulation one searches for the optimum performance by simply following a maximum local ascent. There are two alternatives. \citet{Ince13} followed a bottom-up variant that starts with the best individual neuron and gradually adds neurons. In addition, we also considered a top-down variant that iteratively subtracts neurons starting from the full population. (iii) A conventional albeit heuristic approach taken from statistical physics is simulated annealing. It is known for being less prone to getting stuck in suboptimal solutions as the aforementioned gradient ascent optimization methods.

For the labeled line case, we introduce a novel algorithm for identifying the most discriminative LL population. 
It performs a selection process that evaluates each individual neuron separately and forms the optimized subpopulation by combining the best neurons from every pair of stimuli.

\section{\label{s:Data}Data}

For both the SP and the LL case we used a Poisson neuron model with an absolute refractory period of 2~ms. We always considered pure time coding. All coding and non-coding neurons had the same baseline rate $M$. For our first examples we also assumed the coding of the optimal subpopulation to be noiseless and perfect, though later on this assumption was weakened by adding noise. 

\subsection{\label{ss:SP-data}Summed Population (SP)}

\begin{figure}[!ht]
	\includegraphics[width = 0.48\textwidth]{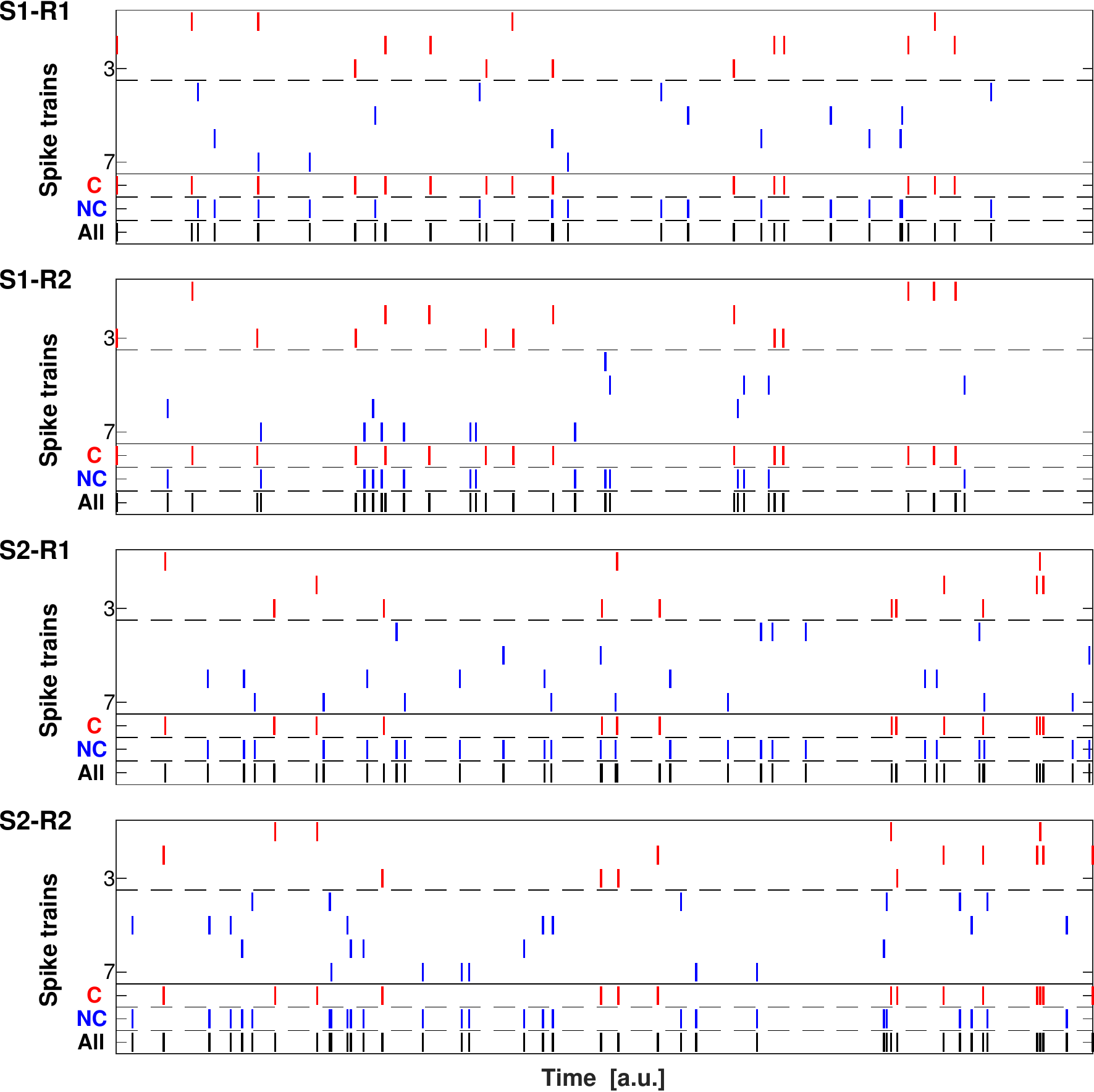}
	\caption{\label{fig:SumPop_Spike_Train_Example} Summed population coding: Simulated spike train responses obtained from the first two of $R\!=\!5$ repetitions for the first two of $S\!=\!4$ stimuli (i.e. overall four out of $T\!=\!20$ trials). There are $N = 7$ neurons of which $c\!=\!3$ form the coding subpopulation (in red) whereas the remaining $4$ neurons are just noisy (in blue). Below the spikes of the individual neurons we depict the pooled spike train of the coding subpopulation ("C", red), the noisy (non-coding) subpopulation ("NC", blue) and the whole population ("All", black). By construction, the pooled spike train of the coding subpopulation was identical for different repetitions of the same stimulus (i.e., for the first two and for the last two rasterplots), while the pooled response of the non-coding subpopulation was random.}
\end{figure}

We simulated spike train responses for a group of neurons that code in unison but not individually. These coding neurons were complemented by non-coding ones, which were simulated separately as neurons without responses beyond baseline activity. We show the generation of the SP data using an example with $S\!=\!4$ stimuli and $R\!=\!5$ repetitions each, so overall $T\!=\!20$ trials. The population comprised of $N\!=\!7$ neurons, the first $c\!=\!3$ of which were coding perfectly for the different stimuli and the last four were non-coding. Fig.~\ref{fig:SumPop_Spike_Train_Example} depicts four exemplary spike train raster plots: responses to the first two repeated presentations of the first two stimuli.
The same data also serve to illustrate the further procedures in Figs. \ref{fig:Bad-good-distance-matrices} and \ref{fig:Gradient_Illustration_TD_BU}.

The pooled spike train of the first $c$ coding spike trains was generated randomly but different for each of the $S$ stimuli. Subsequently, for each of the $R$ trials of every stimulus the spikes of the pooled spike train were evenly distributed among the $c$ coding neurons. For the $N-c$ noisy, non-coding neurons, a trial was independent of every other trial irrespective of the stimulus. Throughout procedures, we ensured that for each trial the expectation value for the rate was the same for all individual neurons. Hence, the SP activity of the coding neurons for a given stimulus agreed exactly across trials but the activity of both coding and non-coding individual neurons remained largely random. As a result, the coding subpopulation discriminated the different stimuli perfectly, while all its 'superpopulations' (populations which contain it as subpopulation) as well as all its subpopulations did not perform likewise well.

\subsection{\label{ss:LL-Data}Labeled Line (LL)}

To study the LL case, we combined a set of stimuli with a population of neurons by varying the responsiveness of the neurons to these stimuli with the following setting in mind: Usually, every stimulus consists of a combination of different features and the individual neurons are either sensitive to these features or not. This implies that a stimulus may be coded by more than one neuron but also that for a diverse set of stimuli a combination of neurons is required to discriminate between all of them. Similarly, every individual neuron may be sensitive to more than one stimulus, to only one or even to none of the stimuli.

When simulating the spike trains, we assumed that for a neuron not sensitive to any of the features present in a stimulus, every stimulus repetition yielded random firing. On the contrary, a neuron sensitive to a certain stimulus responded very reliably (consistently) to repeated presentations of that very stimulus. In essence, we created a single spike train and copied it. In order to make the resulting spike trains more realistic, we used different realizations of jitter noise \highlight{(up to $\pm 5$ ms)} for every repetition. We note that changing the amplitude of the jitter altered the reliability (consistency) of the responses and we were able to control the responsiveness of the neurons to certain stimuli. As in the SP case, we always used a (statistically) constant rate for all spike trains, while controlling for reliability.

%
%

\section{\label{s:Methods}Methods}

Our approach to assess population coding is to search for the subpopulation with maximum discriminative power across the responses to repeated presentations of a set of stimuli. Since the summed population and the labeled line hypotheses make different assumptions about neuronal coding, this task is addressed using fundamentally different algorithms. However, both analyses rely on similar pairwise distance matrices of the SPIKE-distance $D$, which in the case of SP are calculated for neuronal subpopulations, in the case of LL for individual neurons. The two algorithms also share the same basic discrimination performance $P^{s, \bar s}$ which quantifies for all repetitions $R$ of each pair of stimuli ${s, \bar s}$ the degree to which identical stimuli give rise to similar responses and different stimuli result in dissimilar responses:
\begin{equation} \label{eq:Discrimination_Performance}
	\begin{array}{rcl} \displaystyle
	P^{s, \bar s} \hspace*{-.5em} & = & \hspace*{-.5em}\displaystyle \frac{1}{R^2} \sum_{r, \bar r} D^{s,r;\bar s,\bar r} - \frac{1}{2R(R\!-\!1)} \sum_{s,r,\bar r\neq r} D^{s,r;s,\bar r} \\[4ex]
	& = & \hspace*{-.5em}\displaystyle \langle D^{s,r;\bar s,\bar r} \rangle_{r,\bar r} - \langle D^{s,r;s,\bar r} \rangle_{s,r,\bar r\neq r}.
	\end{array}
\end{equation}
Hence, the larger the mean inter-stimuli distance and the smaller the mean intra-stimulus distance, the better the two stimuli can be distinguished.

In the SP case for the pooled spike trains of every subpopulation under evaluation the discrimination performance is computed for all stimulus pairs at the same time. The most discriminative SP subpopulation is the one that yields the highest average performance. In contrast, for LL every individual neuron is evaluated separately. Since often different stimulus pairs will be distinguished best by different neurons, the discrimination performance is optimized for one stimulus pair at a time. For every stimulus pair the algorithm identifies the discriminative neurons and selects the best one. Together, the selected neurons form the most discriminative LL subpopulation.

\subsection{\label{ss:SP-descrimination} Summed population (SP)}

\subsubsection{\label{sss:Discrimination} Discrimination Performance}

The first step of the summed population analysis for any given subpopulation $\K$, stimulus $s$ and repetition $r$, is to pool the spike trains from all the neurons of this subpopulation according to
\begin{equation}
	t_{\K,s,r} = \bigcup_{k\in\K} t_{k,s,r}.
\end{equation}
The matrix of all pairwise spike train distances between all $T = S\!\cdot\!R$ pooled responses can be readily determined. We denote them as
\begin{equation}
	D_\K^{s,r;\bar s,\bar r} = D\left(t_{\K,s,r}, t_{\K,\bar s,\bar r}\right).
\end{equation}

If a neuron population is able to discriminate a pair of stimuli, one can expect low values of the spike train distances between different repetitions of the same stimulus (intra-stimulus), but high values for different stimuli (inter-stimulus). In the SP case all stimulus pairs are evaluated at the same time. Accordingly, one can introduce the discrimination performance of a subpopulation $\K$ as
\begin{equation}\label{eq:SP_Discrimination_Performance}
	\begin{array}{rcl} \displaystyle
	P_\K \hspace*{-.5em} & = & \hspace*{-.5em}\displaystyle \frac{1}{S\!(S\!-\!1)} \sum_{s,\bar s\neq s} P_\K^{s,\bar s}\\[4ex]
	& = & \hspace*{-.5em}\displaystyle \frac{1}{S\!(S\!-\!1)R^2} \! \sum_{s,\bar s\neq s,r,\bar r} \! D_\K^{s,r;\bar s,\bar r} \!-\! \frac{1}{S\!R(R\!-\!1)} \! \sum_{s,r,\bar r\neq r} \! D_\K^{s,r;s,\bar r} \\[4ex]
	& = & \hspace*{-.5em}\displaystyle \langle D_\K^{s,r;\bar s,\bar r} \rangle_{s,\bar s \neq s,r,\bar r} - \langle D_\K^{s,r;s,\bar r} \rangle_{s,r,\bar r\neq r},
	\end{array}
\end{equation}
with $P_\K^{s,\bar s}$ given in Eq.~\ref{eq:Discrimination_Performance} and here specified via subscript to the subpopulation $\K$. The better subpopulation $\K$ is able to distinguish the different stimuli, the higher the value of $P_\K$. Note that in Eq.~\ref{eq:Discrimination_Performance} we could reduce the computational cost by making use of the fact that the initial loop $\langle \text{inter-stimuli} \rangle - \langle\text{intra-stimulus} \rangle$ over stimulus pairs can be transformed into the mean inter-stimulus distance minus the mean intra-stimulus distance for the entire discrimination matrix.

\begin{figure}[!ht]
	\includegraphics[width=\columnwidth]{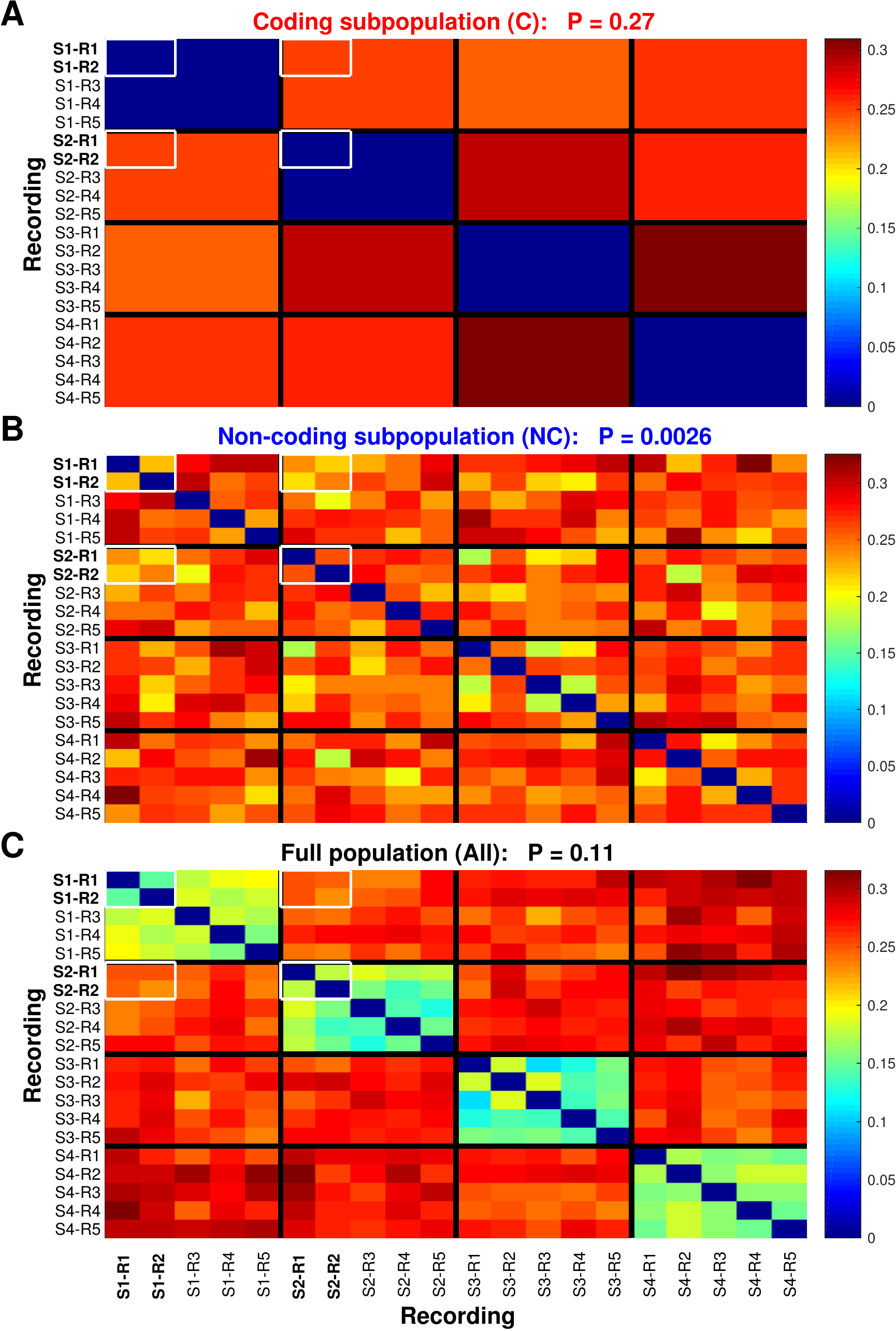}
	\caption{\label{fig:Bad-good-distance-matrices}
Summed population coding: Stimulus-dependent clustering for the seven neurons (three coding, four non-coding) of Fig.~\ref{fig:SumPop_Spike_Train_Example}:
Pairwise spike train distance matrices of all $T\!=\!20$ trials consisting of $S\!=\!4$ stimuli with $R\!=\!5$ repetitions each for three different subpopulations:
A. The coding subpopulation consisting of first three neurons (C, red) distinguish the stimulus perfectly. The different stimuli can be distinguished easily because in this noise-free case high distances are obtained for inter-stimuli realizations and zero values for intra-stimuli realizations. Accordingly a very large discrimination performance is obtained.
B. Evaluating the summed activity of the last four neurons, the non-coding subpopulation (NC, blue) leads to seemingly random distances and stimulus discrimination fails. Such a distance matrix results in a very low discrimination performance.
C. For the full population (All, black) the intra-stimulus sub-matrices can still be distinguished but are much less pronounced. Accordingly, the discrimination performance $P$ attains some intermediate value. The sub-matrices resulting from the four examples (first two repetitions of the first two stimuli) given in Fig.~\ref{fig:SumPop_Spike_Train_Example} are marked by white boxes.}
\end{figure}

Fig.~\ref{fig:Bad-good-distance-matrices} shows the pairwise distance matrices and their discrimination performance values for three subpopulations of the data set exemplarily shown in Fig.~\ref{fig:SumPop_Spike_Train_Example}. In this noise-free example, the coding subpopulation (the first three neurons in Fig.~\ref{fig:Bad-good-distance-matrices}A) was able to discriminate perfectly and, accordingly, we obtained a very high value of $P$. The non-coding subpopulation (last four neurons in Fig.~\ref{fig:Bad-good-distance-matrices}A) could not distinguish between the different stimuli. Its discrimination performance $P$ was very close to the expected zero value (Fig.~\ref{fig:Bad-good-distance-matrices}B). Finally, for the pairwise distance matrix of the full population (Fig.~\ref{fig:Bad-good-distance-matrices}C), which contained both the coding and the noisy non-coding subpopulation, we found an intermediate discrimination performance $P$.

\subsubsection{\label{sss:SP_Algorithms} Algorithms}

Since the measure $P_\K$ in Eq.~\ref{eq:SP_Discrimination_Performance} quantifies the discrimination performance for every given subpopulation, it can serve to search the space of all possible subpopulations for the best SP-coding subpopulation~$\K_\text{opt}$, defined as
\begin{equation}\label{eq:SP_Optimized_Population}
	\K_\text{opt}: P_{\K_\text{opt}}^\text{SP} = \max_\K\,\left\{P_\K\right\}.
\end{equation}
As mentioned in Section \ref{s:SPIKE}, there are three different approaches to this task.

\paragraph{\label{sssp:SP-Algo-Brute_Force}(i) Brute force}
One determine the stimulus discrimination performance $P_\K$ for the summed activity of every possible subpopulation $\K$, and identifies the subpopulation that provides the maximum performance. Since all possible subpopulations are evaluated, the brute force approach is guaranteed to find the best subpopulation. Its result can thus serve as a ground truth for other less exhaustive algorithms. Evaluating all possible subpopulations, however, is not feasible for very large datasets because the number of possible subpopulations increases exponentially with the number of neurons~$N$:
\begin{equation} \label{eq:Total-number-of-combinations}
	K_\text{bf} = \sum_{k=1}^N \binom{N}{k} = 2^N-1.
\end{equation}
For example, for $N\!=\!125$ the individual terms from different subpopulation sizes add up to $K_\text{bf}\!=\!4.25 \cdot 10^{37}$, which is far beyond the limits of current soft- and hardware implementations. More restrictive algorithms are needed that explore only a (relevant) subspace of the numerous subpopulations.

\paragraph{\label{sssp:SP-Algo-Gradient}(ii) Gradient algorithms}
The idea behind gradient algorithms is to evaluate the discrimination performance for a restricted number of neuronal subpopulations. There are two variants:
The {\em{bottom-up variant}} used by \citet{Ince13} starts with the best individual neuron and builds up the population by adding in each iteration the best remaining neuron. The alternative {\em{top-down variant}} starts from the complete population and iteratively subtracts one neuron at a time.
\begin{figure}[!ht]
	\includegraphics[width=\columnwidth]{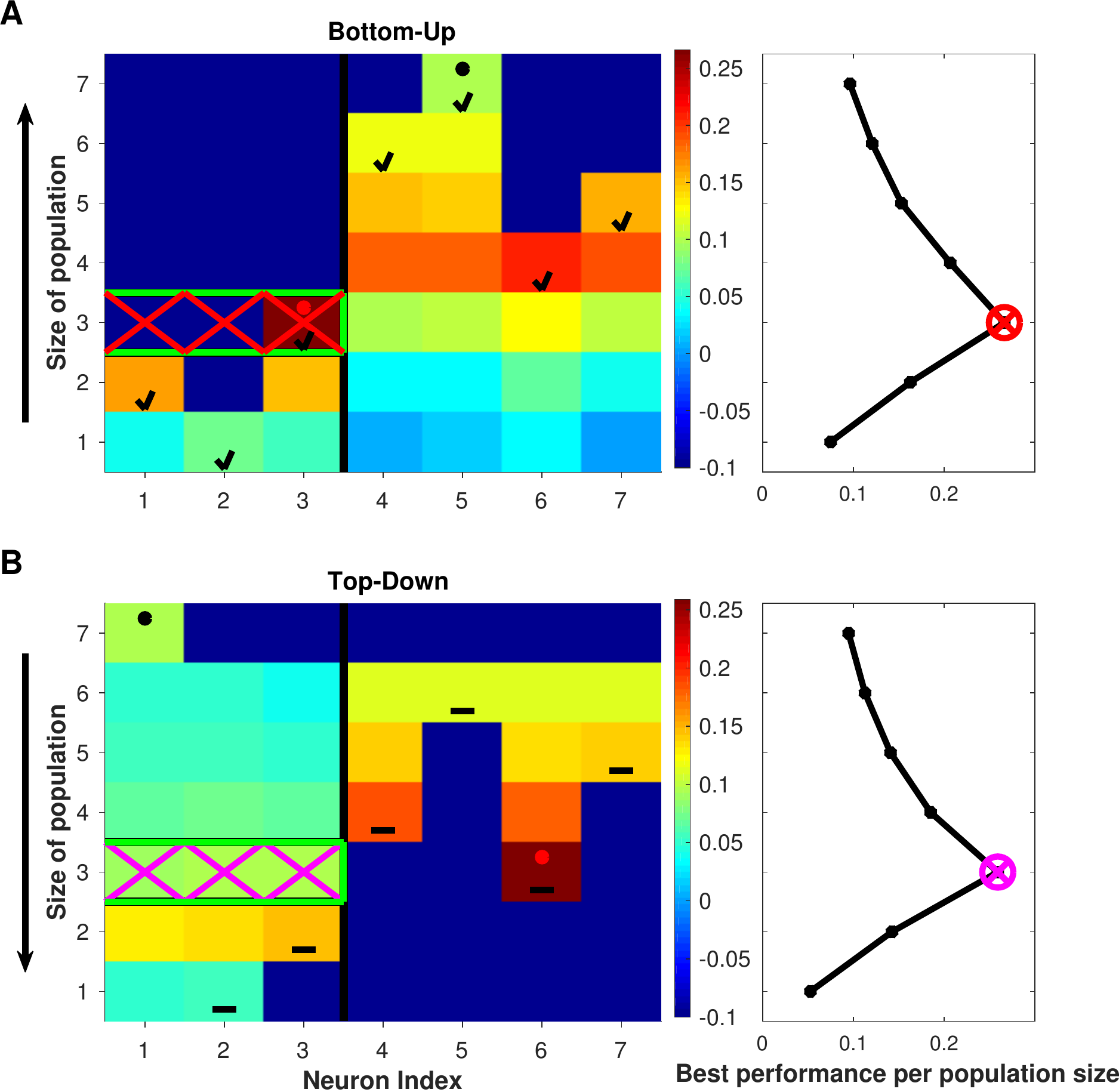}
	\caption{\label{fig:Gradient_Illustration_TD_BU} Summed population coding: Color-coded discrimination performance for different neuronal subpopulations within the example from Figs. \ref{fig:SumPop_Spike_Train_Example} and \ref{fig:Bad-good-distance-matrices}, in which a subpopulation consisting of the first $3$ out of $N = 7$ neurons code for the different stimuli while the remaining non-coding neurons fire just randomly. 
Every matrix element depicts the performance of one specific subpopulation.
The right panels show the maximum performance for a given subpopulation size.
A. The bottom-up algorithm starts with the discrimination performances of the individual neurons.
In every subsequent iteration one adds the neuron that complements the current subpopulation best (indicated by small black ticks) and this is repeated until the full population size is reached.
B. The top-down algorithm begins with the discrimination performance of the complete population (depicted in the first column of the top row).
In every iteration, one discards the neuron that contributes the least to the discrimination (marked by short black horizontal lines) until just one neuron remains. ---
The red and the black dots in both matrices mark the coding and the full subpopulations whose pairwise distance matrices are shown in Fig.~\ref{fig:Bad-good-distance-matrices}A and \ref{fig:Bad-good-distance-matrices}C, respectively (the non-coding subpopulation in Fig.~\ref{fig:Bad-good-distance-matrices}B is never visited).
For both algorithms, the maximum overall population sizes (red circle in A, \highlight{magenta} circle in B) is correctly obtained for the coding subpopulation (marked by crosses in red resp. \highlight{magenta} and confirmed by the green rectangles indicating the ground truth results of the brute force approach).}
\end{figure}
Both variants are illustrated in Fig.~\ref{fig:Gradient_Illustration_TD_BU} using the example from Figs. \ref{fig:SumPop_Spike_Train_Example} and \ref{fig:Bad-good-distance-matrices}. 
In this example, both gradient variants correctly identified the first $3$ neurons as the coding subpopulation. Importantly, for either algorithm the number of combinations for which the stimulus discrimination performance had to be calculated amounts only to
\begin{equation} \label{eq:Selected-number-of-combinations}
	K_\text{grad} = \sum_{k=1}^N k = \tfrac{1}{2}{N(N+1)},
\end{equation}
which is much smaller than $K_\text{bf}$ and, thus, feasible even for very large $N$. For $N = 125$, the individual terms from different subpopulation sizes add up to only $K_\text{grad}=7875$.

\paragraph{\label{sssp:SP-Algo-SimAnn}(iii) Simulated annealing}
Simulated annealing \citep{Dowsland12} is a heuristic approach, which -- in principle -- allows to find the global maximum without having to explore the whole search space, though there is no guarantee that the optimum solution will at all be found and, if so, that this will be in fewer steps than the brute force search. However, simulated annealing, in contrast to the two gradient algorithms, has the ability to recover from local maxima. Suboptimal solutions are hence much less likely to occur.

One uses a random permutation of neurons as an initial subpopulation $\K_0$. The $n$-th step in the search is to add or remove a randomly chosen neuron to or from the current subpopulation $\K_{n-1}$ resulting in a new population $\K_n$. Addition or removal is applied with equal probability, except for the boundary populations of one neuron and all neurons, for which the only possible steps are to add or to remove a neuron, respectively. Whether or not the addition/removal is accepted depends on the new discrimination performance $P_n$ relative to the current one $P_{n-1}$. The corresponding acceptance probability is set by
\begin{equation} \label{eq:acceptance_probability}
q_n = \exp \left\{-\frac{\left|P_{n}-P_{n-1}\right|}{\mathcal{T}_{n-1}}\right\},
\end{equation}
where $\mathcal{T}_{n-1}$ is a pseudo-temperature that allows moving 'downhill' in order to not get stuck in a local and thus suboptimal maximum. Steps with $P_{n}>P_{n-1}$ are always accepted. The likelihood of accepting steps with $P_{n}\leq P_{n-1}$ is also finite but decreases according to a gradual and stepwise cooling scheme in which $\mathcal{T}_n$ is held constant for a certain number of iterations chosen depending on the number of neurons.

By means of a path of $N_0$ random test steps from the starting population one can set the initial temperature to
\begin{equation} \label{eq:initial_temperature}
\mathcal{T}_0 = 
-\tfrac{1}{\ln(0.95)}\left\langle\left|P_{n}-P_{n-1}\right|\right\rangle_{n=1,\dots,N_0},
\end{equation}
which guarantees a fair mobility in the beginning because even downhill steps are accepted with a likelihood of $95\%$ \citep{BenAmeur04}. The stopping criterion for the algorithm is that between two successive temperature changes the population remains unchanged, i.e. $\K_{n}=\K_{n-1}$. During the whole iteration one tracks the highest discrimination performance value reached thus far and in case the final value is worse than this best value along the path, it can not be the global maximum and thus the algorithm resets the temperature to $\mathcal{T}_0$ and continues with increased mobility.

\subsection{\label{ss:LL-Methods}Labeled Line (LL) -- Discrimination Performance \& Algorithm}

The assumption underlying LL coding is that neurons individually encode different properties or features of a stimulus. Hence, in this case every neuron must be evaluated separately. This actually makes things much easier since instead of having to deal with distance matrices in the space of all possible subpopulations as in the SP case, it is now sufficient to calculate only $N$ distance matrices, one for each neuron. A single neuron typically codes for one specific stimulus feature only. Therefore, in order to discriminate a large and broad set of $S$ stimuli, the complementary information provided by many neurons needs to be combined. Identifying the most discriminative LL population is thus equivalent to finding discriminative neurons for as many stimulus pairs as possible. For two stimuli to be distinguishable there must be at least one individual neuron sensitive to their difference. In case more than one neuron is found for a stimulus pair, the most discriminative one is selected.

\begin{figure*}[t]
	\includegraphics[width = 1.0\textwidth]{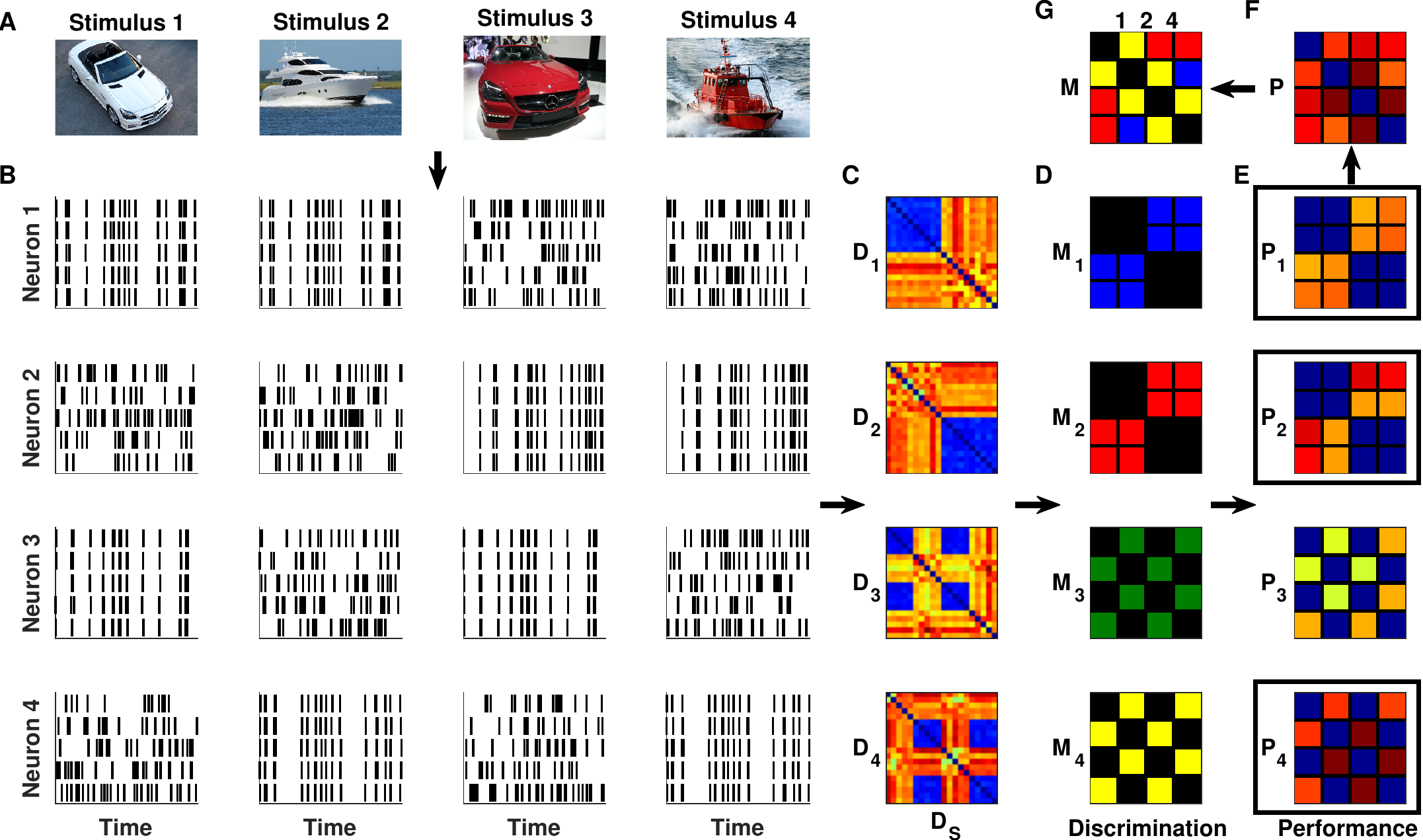}
	\centering
	\caption{\label{fig:LL_population}
Labeled line coding:	Schematic example in which each of the $N = 4$ neurons is sensitive only to one specific feature of the different stimuli. Neuron 1 responds to white and neuron 2 to red objects, neuron 3 to cars and neuron 4 to ships. A. The $S = 4$ stimuli were chosen such that they combine these features. B. Spike trains responses of each neuron to $R = 5$ repetitions of every stimulus. C. Pairwise distance matrices $D_n$ (using the SPIKE-distance $D$) over all $T = 20$ trials. D. Corresponding discrimination matrices $M_n$. Black is always $0$, whereas each neuron has its own color representing $1$ for the stimulus pairs it can distinguish. E. Performance matrices $P_n$. The value is zero for stimulus pairs that can not be discriminated and otherwise the higher the better the discrimination. F. The population performance matrix $P$ collects for each stimulus pair the highest values obtained for any of the four individual neurons. G. The corresponding neurons are indicated in the population discrimination matrix $M$ which is colored according to which neuron achieves the best discrimination performance ($P^{s, \bar s}_n$) for that stimulus pair. The overall performance is $P^{LL} = 0.148$ and is obtained for the optimized LL population $\K_\text{opt}^\text{LL} = [1 \hspace{0.07cm} 2\hspace{0.08cm} 4]$ (written on top, the corresponding performance matrices are marked in E by black rectangles). Note that the color coding in both subplots D and G is discrete and used to label the individual neurons.}
\end{figure*}
Fig.~\ref{fig:LL_population} illustrates this procedure using a very schematic and simplified example. For clarity we use two very distinct features (color and vehicle type), but in a typical recording these often would be two different features of the same sensory mode. Our starting point is a set of $S$ different stimuli (Fig.~\ref{fig:LL_population}A) whose repeated presentations elicit $T = S \times R$ spike train responses (Fig.~\ref{fig:LL_population}B). From these responses a pairwise distance matrix $D$ is computed for every individual neuron $n = 1,...,N$ (Fig.~\ref{fig:LL_population}C). Looping over stimulus pairs transforms each of these $T \times T$ distance matrices into a $S \times S$ discrimination matrix that indicates the stimulus pairs this particular neuron is able to discriminate. For two different stimuli $s$ and $\bar s$ to be distinguishable, their intra-stimulus distances $D^{s,s}$ and $D^{\bar s,\bar s}$ and their inter-stimuli distances $D^{s,\bar s}$ (both pooled over all repetitions of the respective stimuli) should stem from different distributions. If they were to stem from the same distribution, the two stimuli could not be discriminated.

We seek to verify the hypothesis that responses to two different stimuli can be discriminated. To this end, we employ three \highlight{Wilcoxon rank-sum tests} $t_{s,s}^{\bar s,\bar s}=t\left(D^{s,s},D^{\bar s,\bar s}\right)$, $t_{s,s}^{s,\bar s}=t\left(D^{s,s},D^{s,\bar s}\right)$, and $t_{\bar s,\bar s}^{s,\bar s}=t\left(D^{\bar s,\bar s},D^{s,\bar s}\right)$ at a significance level of $\alpha = 0.001$. From these three tests one can form a logical discrimination matrix such that for neuron $n$ the discrimination between stimuli $s$ and $\bar s$ reads
\begin{equation} \label{eq:LL_Discrimination}
	M^{s,\bar s}_n=
	\begin{cases}
		1 & \text{if} \ \ t_{s,s}^{\bar s,\bar s} \, \lor\, t_{s,s}^{s,\bar s} \,\lor\, t_{\bar s,\bar s}^{s,\bar s}~=~true  \\
		0 & \text{otherwise.}
	\end{cases}
\end{equation}
The two stimuli can be discriminated by this neuron (i.e. $M_{s,\bar s}^n = 1$) whenever at least one of the three tests yields significant differences. In the example of Fig.~\ref{fig:LL_population}, all neurons respond to just one single feature of the stimuli, the first two to color (white and red) and the last two to vehicle type (car and ship). Thus, all of the neurons are able to discriminate among some of the stimuli but not among others (Fig.~\ref{fig:LL_population}D).

Next, in order to identify the best LL subpopulation for discrimination, we define the discrimination performance for each stimulus pair $(s, \bar s)$ and every neuron $n$ as
\begin{equation}\label{eq:LL_Discrimination_Performance}
\hspace*{-.7em}
\hat{P}^{s, \bar s}_n =  M^{s,\bar s}_n P^{s, \bar s}_n
\end{equation}
with $P^{s, \bar s}_n$ given in Eq.~\ref{eq:Discrimination_Performance}, supplemented by the subscript $n$ to index individual neurons.
High values of $\hat{P}^{s, \bar s}_n$ are obtained for large inter-stimuli and small intra-stimulus distances, while for the stimuli pairs a neuron cannot discriminate the value vanishes (cf. Fig.~\ref{fig:LL_population}E). 
From these individual discrimination performance matrices the population performance matrix can be obtained as
\begin{equation}\label{eq:LL_Population_Performance}
 	\hat{P}^{s, \bar s}_\text{max} = \max_n \left\{\hat{P}^{s, \bar s}_n\right\}  , 
\end{equation}
which takes for every stimulus pair the best discrimination performance of all the individual neurons (see Fig.~\ref{fig:LL_population}F).
The population discrimination matrix
\begin{equation}\label{eq:LL_Population_Discrimination}
 	M^{s, \bar s}_\text{max} =
 			\begin{cases}
			\arg \hat{P}^{s, \bar s}_\text{max}   & \displaystyle\text{if} \ \ \hat{P}^{s, \bar s}_\text{max} > 0 \\
			0 & \text{\rm otherwise}
		\end{cases}
\end{equation}
indicates for every stimulus pair the best neuron (cf. Fig.~\ref{fig:LL_population}G) and from this matrix the optimized LL population is obtained by uniting all neurons that contribute to the discrimination, i.e.
\begin{equation}\label{eq:LL_Optimized_Population}
	\K_\text{opt}^\text{LL} = \bigcup_{s \neq \bar s, M^{s, \bar s}_\text{max}>0}  M^{s, \bar s}_\text{max}.
\end{equation}
Finally, the LL discrimination performance of the full population for the whole stimulus set is the mean of the  discrimination performances over all stimulus pairs, that is, 
\begin{equation}\label{eq:LL_Performance_value}
 	P^\text{LL} = \left\langle  {\hat{P}^{s, \bar s}_\text{max}}  \right\rangle_{s \neq \bar s}.
\end{equation}
In our example, from Fig.~\ref{fig:LL_population}G we can extract the best selection from the two color neurons (both neuron 1 and neuron 2) and from the two 'vehicle type' neurons (just the ship neuron, neuron 4) yielding neurons 1, 2, and 4 as the optimized LL population which obtains an labeled line discrimination performance of $P^\text{LL} = 0.148$.

%
%

\section{\label{s:Results}Results}

\subsection{\label{ss:SP-Results}Summed Population (SP)}

\paragraph{(i) Brute force}

This is the preferred algorithm because it guarantees that the best discrimination performance is found.
However, the number of subpopulations that have to be evaluated increases exponentially with the number of neurons, see Eq.~\ref{eq:Total-number-of-combinations}, rendering this algorithm applicable only for comparably small numbers of neurons. In this study whenever possible we used it as benchmark to verify the correctness of the solutions found by the other algorithms and to evaluate their decrease in computational cost. Being able to obtain the ground truth this way was most important for the examples with noisy spike trains where the actual result was not known beforehand.
\begin{figure}[!ht]
	\includegraphics[width=\columnwidth]{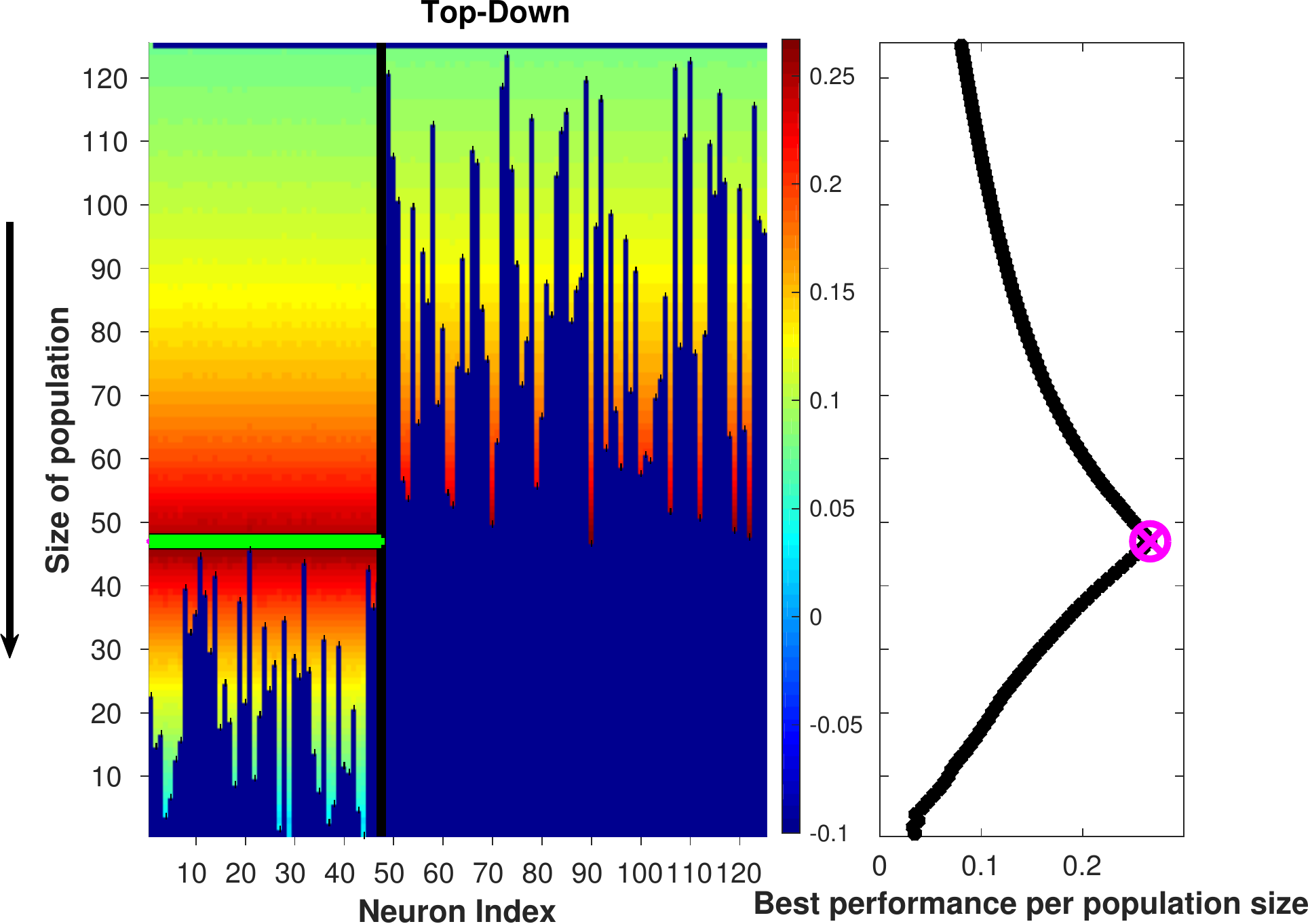}
	\caption{\label{fig:Popp-Perf-125-47} Summed population coding: Same setup and format as Fig.~\ref{fig:Gradient_Illustration_TD_BU}B but this time the top-down gradient algorithm was applied to a neuronal population of size $N = 125$ which corresponded to a total number of more than $10^{37}$ possible subpopulations. In this simulation the subpopulation consisting of the first $47$ neurons (marked in green) coded for the different stimuli while the remaining non-coding neurons fired randomly. Even though it evaluated just less than $8.000$ subpopulations, the algorithm correctly identified the coding subpopulation as indicated by the maximum over all population sizes (\highlight{magenta} circle in curve on the right).}
\end{figure}

\paragraph{(ii) Gradient algorithms}

We illustrate the appropriateness of using gradient algorithms, i.e a proof-of-principle, using a noise-free case.
In Fig.~\ref{fig:Popp-Perf-125-47} we used a similar example as in Fig.~\ref{fig:Gradient_Illustration_TD_BU}, again 
with $S\!=\!4$ different stimuli which were repeated $R\!=\!5$ times each. This time, however, the population consisted of $N = 125$ neurons, a number in the range of real life experiments. This corresponds to more than $K_\text{bf}\!=\!4 \!\cdot\! 10^{37}$ possible subpopulations and thus renders the brute force approach absolutely unfeasible. For the gradient algorithm (top-down variant) the discrimination performance had to be determined for only $7875$ subpopulations. The first $c\!=\!47$ coded perfectly and this coding subpopulation was correctly identified as the one with the maximum discrimination performance $P_{\K_\text{opt}}^\text{SP}$.

Next, we ran two simulations that are constructed such that each time one of the two variants of the gradient algorithm did not find the best subpopulation since it got trapped in a local maximum. \highlight{The two cases employed essentially the same setup, but the first case was noiseless whereas in the second case we applied noise that disrupted the timing information of each spike with 50\% chance.} For the first simulation we generated a population of $N=10$ neurons made up of three different subpopulations. The first four individual neurons are each able to discriminate between all the stimuli on their own. Next, a subpopulation of three neurons could discriminate only collectively, i.e. as a population, but with a slightly lower discrimination than the individual neurons. Finally, the last three neurons were non-coding. For a population of this size the brute force approach is still feasible and its result served as ground truth.

\begin{figure}[!ht]
	\includegraphics[width = \columnwidth]{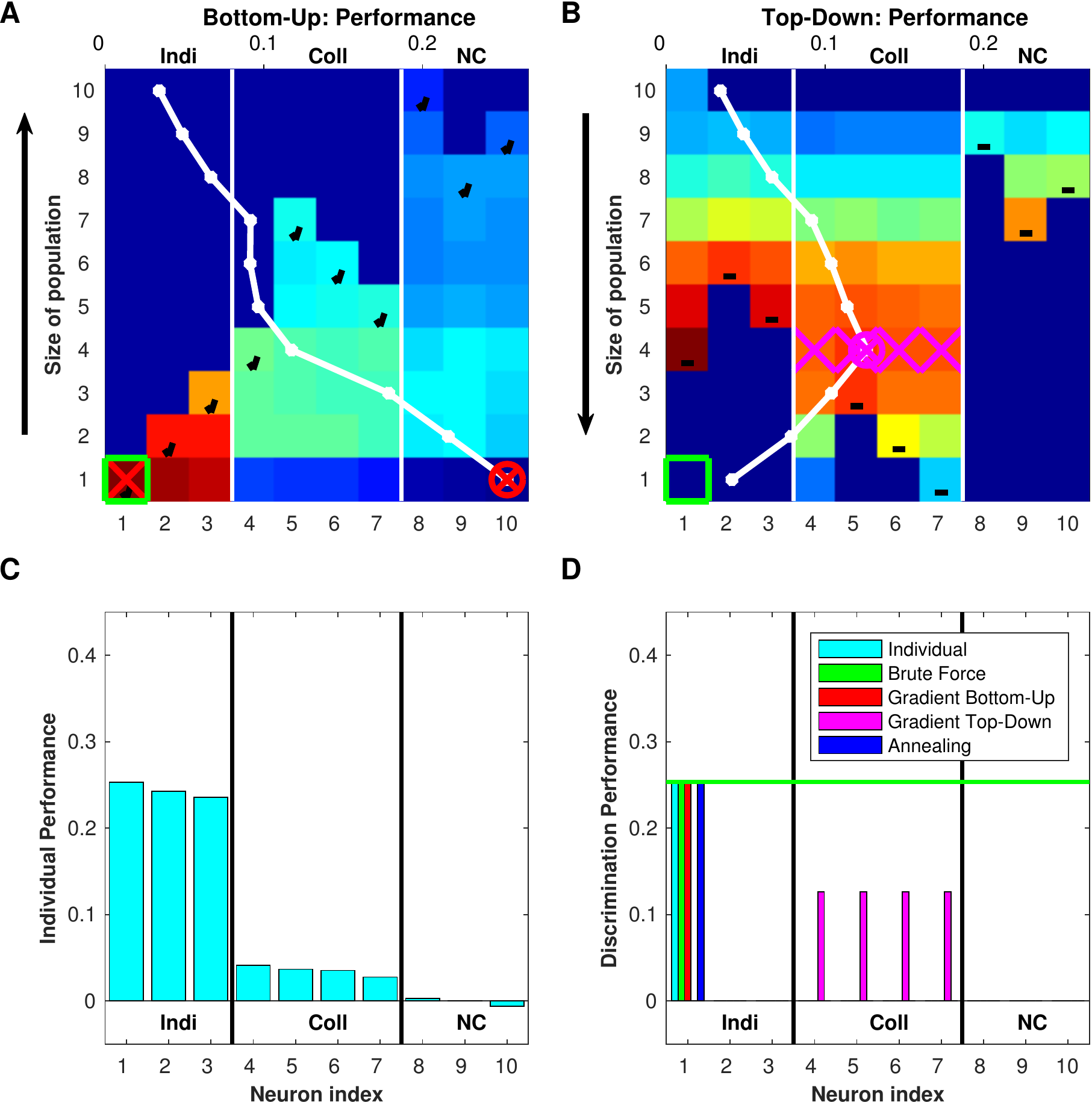}
	\caption{\label{fig:Gradient_TopDown_Fail} Summed population coding: Example where the top-down gradient algorithm failed.
The population consists of three neurons that code the different stimuli individually (Indi), four neurons that code them as a collective (Coll) and three non-coding neurons (NC).
The top panels A and B follow the setup of the bottom-up and the top-down algorithms in Fig.~\ref{fig:Gradient_Illustration_TD_BU}A and B, respectively.
For the sake of legibility now the curves of the maximum performance per population size are superimposed in white (axis on top).
The optimal solution is the single best individually coding neuron (neuron $\#1$ on the left) and this neuron was indeed correctly identified by all algorithms apart from one.
The top-down algorithm always discarded the neuron contributing the least and since it could not discard the whole collective unit it got stuck in a local maximum.
C. Performance of individual neurons.
D. Neuronal subpopulations identified as winners by the different algorithms. The horizontal green line indicates the optimal discrimination performance as verified by the brute force algorithm.}
\end{figure}

In the simulation without noise shown in Fig.~\ref{fig:Gradient_TopDown_Fail}, the best discrimination performance should have been obtained by the very best individual neuron. This was indeed the result of the bottom-up variant (Fig.~\ref{fig:Gradient_TopDown_Fail}A). However, the top-down variant failed and erroneously indicated the neuronal subpopulation consisting of the middle four neurons as the winner (Fig.~\ref{fig:Gradient_TopDown_Fail}B).
This is because it had to follow the iterative procedure of singling out one neuron at a time for elimination and, hence, could not treat the middle population as a single entity. Therefore, at each step breaking up the population was being considered a very bad option and falsely avoided to the very end. The individually coding neurons were eliminated first and thus never evaluated on their own, which left the performance of the collectively coding subpopulation as the best one encountered along the path.

Fig.~\ref{fig:Gradient_TopDown_Fail}C depicts the SP discrimination performances given in Eq.~\ref{eq:SP_Discrimination_Performance} calculated for each of the individual neurons. Each of the first three neurons had a very large discriminative power far superior to the individual neurons of the collectively coding subpopulation which were still better than the non-coding neurons. In Fig.~\ref{fig:Gradient_TopDown_Fail}D we show the winners chosen by each of the different algorithms. The top-down gradient variant was the only algorithm that did not succeed in identifying the very first individual neuron as the most discriminative subpopulation (as verified by the brute force approach).

\begin{figure}[!ht]
	\includegraphics[width = \columnwidth]{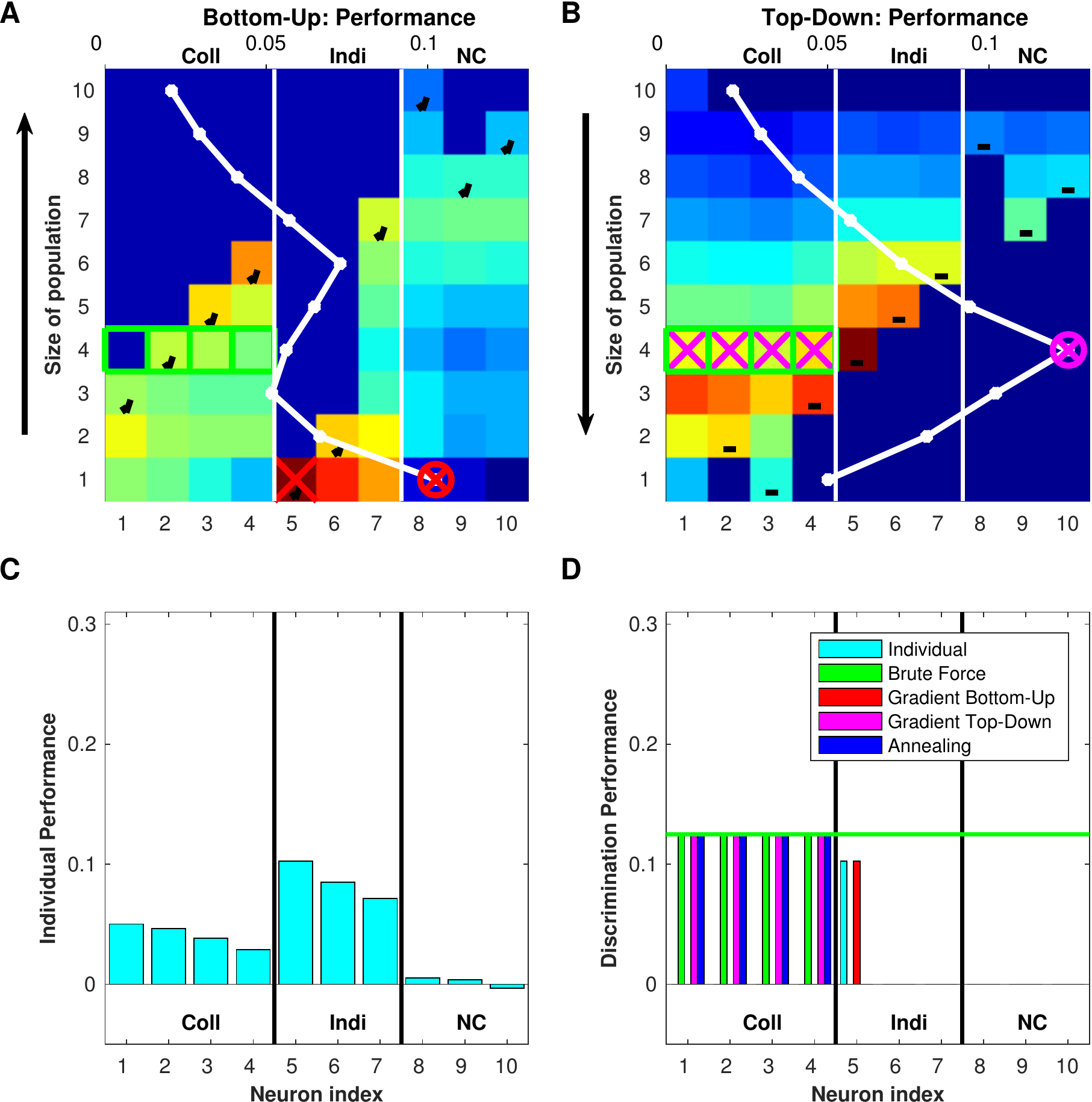}
	\caption{\label{fig:Gradient_BottomUp_Fail} Summed population coding: Similar to Fig.~\ref{fig:Gradient_TopDown_Fail} but in this example we added some noise which made the bottom-up gradient algorithm fail. The discriminative performance of the individual neurons was degraded to such an extent that here the four summed population neurons performed better (to stay with our convention that the winning subpopulation comes always first, we reversed the order of these two groups of neurons).
The bottom-up algorithm could never add the SP-population as a whole and thus always had to pick the best individual neuron remaining.
This took it to a local maximum right at the beginning, and from then on the discrimination performance of the very first individual neuron apparently remained the optimal solution along the whole path.}
\end{figure}
In the second simulation summarized in Fig.~\ref{fig:Gradient_BottomUp_Fail}, we used exactly the same setup but added so much noise to the first three individually coding neurons that they did no longer outperform the four collectively coding neurons which thus together should have been identified as the most discriminative subpopulation. In this case the bottom-up gradient algorithm failed (Fig.~\ref{fig:Gradient_BottomUp_Fail}A), whereas the top-down variant managed to find the correct solution (Fig.~\ref{fig:Gradient_BottomUp_Fail}B). The reasoning is exactly inverse to the first case. The iterative 'one-neuron at a time' scheme did not allow to include the collectively coding population consisting of four neurons in one step and since each of them on their own was worse than the three individually coding neurons (see Fig.~\ref{fig:Gradient_BottomUp_Fail}C), these latter neurons were added first. Again, Fig.~\ref{fig:Gradient_BottomUp_Fail}D summarizes the results of the different algorithms. There, the bottom-up gradient algorithm incorrectly identified the best individual neuron as most discriminative. 

The respective failures of the top-down and the bottom-up variants in these two simulations were both due to the fact that the gradient algorithms follow a steepest ascent approach where at each step they take the locally optimal choice. The well-known problem with this \textit{greedy} approach is that it does not necessarily lead to the global optimum. In this specific context it meant that once an incorrect neuron was added (bottom-up) / a correct neuron was excluded (top-down), the gradient algorithms had no way to correct this 'mistake' and these bad choices remained. 
We would like to add that we also found cases in which neither variant was able to find the correct solution. So running both algorithms and picking the overall best performance can also not provide a guarantee that the optimal solution is found.

\paragraph{(iii) Simulated annealing}

Since gradient algorithms are much faster than the brute force approach and successful under idealized conditions, they can be used for first testing. However, our examples illustrate that they can generally not be relied upon in more realistic settings. Fortunately, simulated annealing provides a recovery mechanism that considerably reduces the likelihood of getting stuck in a local maximum.

Both Fig.~\ref{fig:Gradient_TopDown_Fail}D and Fig.~\ref{fig:Gradient_BottomUp_Fail}D include the successful simulated annealing algorithm. In these two cases as well as in all other cases that we looked at, simulated annealing was indeed able to identify the most discriminative neuronal subpopulation. However, this increased reliability of the simulated annealing algorithm compared to the two gradient variants comes with a prize, an increased computational cost.

\begin{figure}[!ht]
\centerline{\includegraphics[width=\columnwidth]{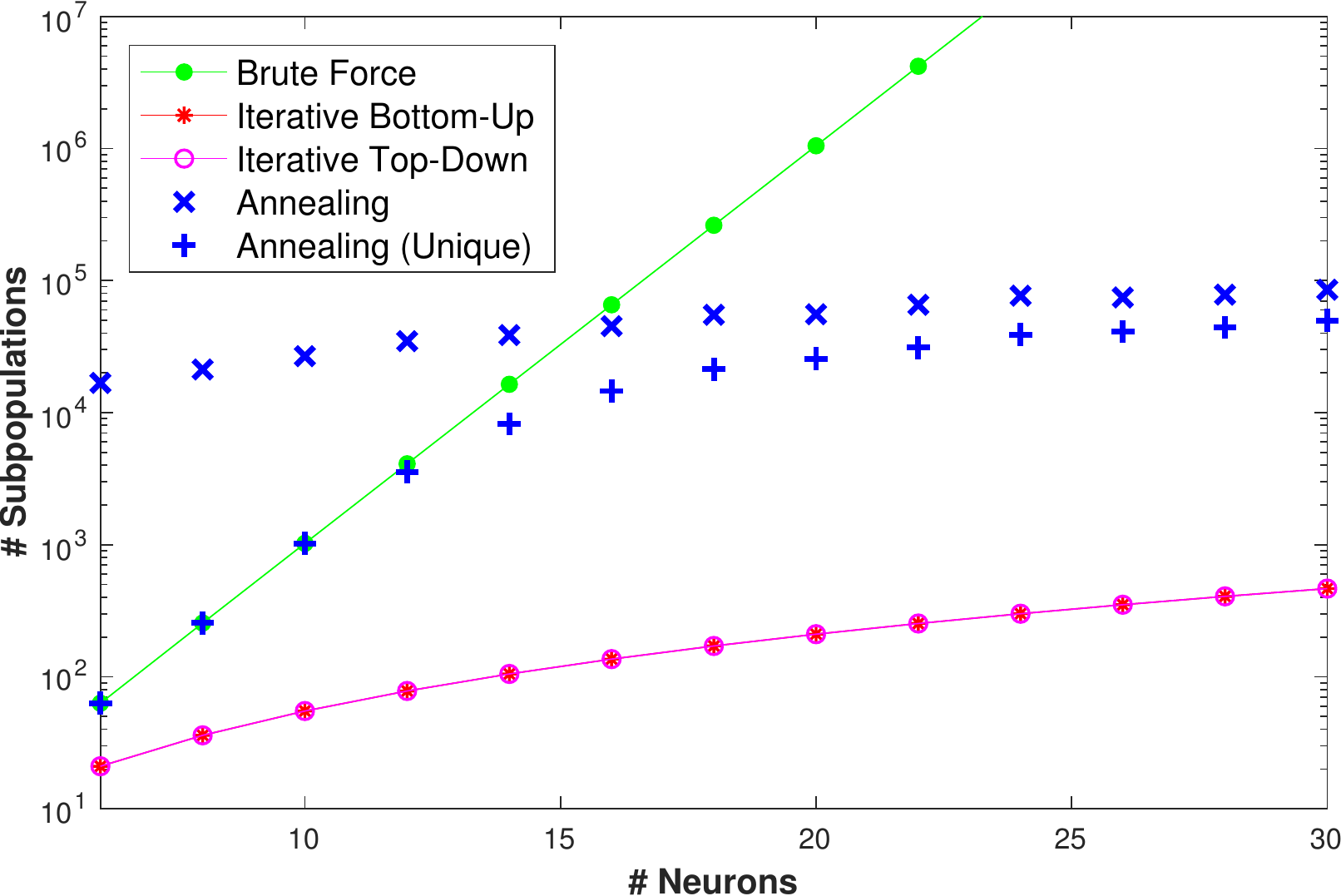}}
	\caption{\label{fig:Speed}
Comparison of the computational cost for the different algorithms for summed population coding: Subpopulations evaluated versus number of neurons. \highlight{Note that the values for simulated annealing are averages over ten trials.}}
\end{figure}
The runtime of an algorithm consists mostly of two parts, the number of subpopulations that have to be evaluated and the time it takes to evaluate each of these subpopulations. As summarized in Fig.~\ref{fig:Speed}, we compared the different algorithms regarding the numbers of subpopulations visited. As expected, for the brute force algorithm that number increased exponentially with the number of neurons. For the two variants of the gradient algorithm the numbers were identical in line with Eq.~\ref{eq:Selected-number-of-combinations}, and much smaller than for the brute force algorithm. 
In the case of simulated annealing, our selected examples revealed values in between these two extremes for large enough numbers of neurons.
Here we had to divide the number of subpopulations evaluated into two distinct categories. The first one is the actual number visited by the random walk, the second one is the number of uniquely evaluated subpopulations. We added the latter because some of the subpopulations might have been revisited many times. We note that in our implementation the discrimination performance for every subpopulation is determined only at the first visit and stored so that at all repeated visits this value can be readily retrieved via an unambiguous mapping to the subpopulation space. This makes revisiting a subpopulation considerably cheaper than calculating a new one. While for small population sizes the actual number of subpopulations evaluated might be even higher than for brute force, it starts to gain more speed compared to the brute force approach as soon as the algorithm no longer has to visit all solutions. This also implies that for a small number of neurons it is always preferable to use the brute force approach, because it not only gains speed, but, unlike the simulated annealing, also guarantees the ground truth solution.

\subsection{\label{ss:LL-Results}Labeled line (LL)}

\begin{figure*}[ht!]
	\includegraphics[width = 1.0\textwidth]{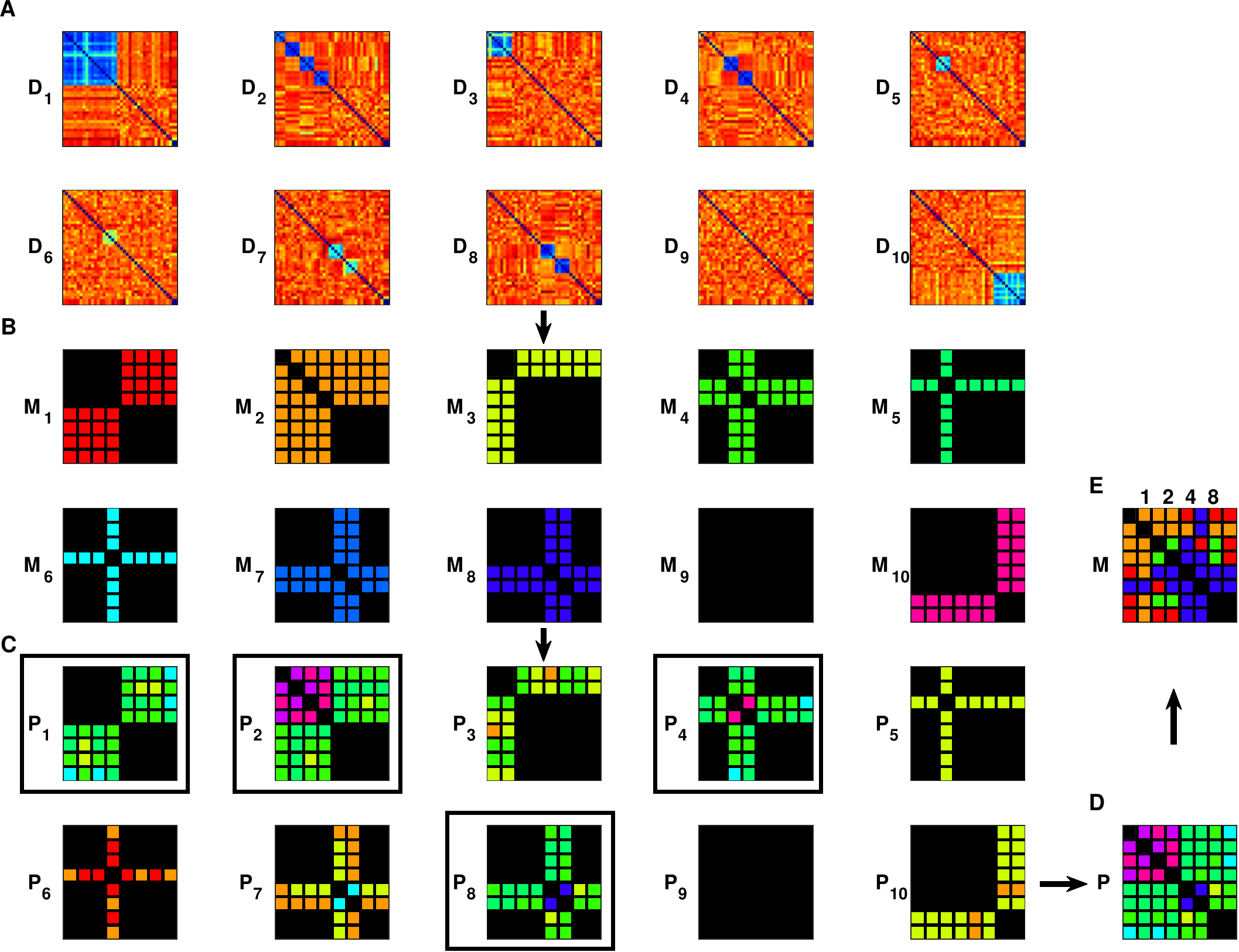}
	\centering
	\caption{\label{fig:LL_population_complex} Labeled Line Coding: A reasonably complex example based on $R\!=\!5$ repetitions, $S\!=\!8$ stimuli and $N\!=\!10$ neurons.
A. Pairwise spike train distance matrices $D_n$ over all $T\!=\!40$ trials.
Both neurons 1 and 2 responded to the first four stimuli, but whereas the first reacted with just one common response, the second did so with a different response for each of these stimuli. Neuron 3 was only sensitive to the first two (equally) and neuron 4 only to the third and fourth stimulus (separately) who were also covered by neuron 5 and 6 but only one at a time. The two neurons 7 and 8 coded for the fifth and sixth stimulus but with varying degrees of reliability, while \highlight{neuron 9} did not respond to any of the stimuli. \highlight{Neuron 10} was sensitive to the last two stimuli reacting to both of them with the same response.
B. Discrimination matrices $M_n$,  colored such that the stimuli pairs that could be discriminated are marked with a unique color for each neuron.
C. Corresponding performance matrices $P_n$ that were used to identify the best performance for each stimulus pair.
D. Maximization of the $P_n$-matrices results in the population performance matrix $P$.
The black off-diagonal element indicates stimuli 7 and 8 which could not be discriminated by any of the neurons of this population.
E. The population discrimination matrix $M$ gathers the neurons that contributed to the optimized LL subpopulation $\K_\text{opt}^\text{LL} =  [1 \hspace{0.08cm} 2 \hspace{0.08cm} 4 \hspace{0.08cm} 8]$ (again marked in C by black rectangles). Their overall performance in this example was $P_{LL} = 0.139$.}
\end{figure*}

For the LL case, we investigated a more complex example than Fig.~\ref{fig:LL_population}. However, we used a schematic design to sketch a general idea of some of the complications that may occur. In this example $R\!=\!5$ repetitions of $S\!=\!8$ stimuli were presented to $N = 10$ neurons (Fig.~\ref{fig:LL_population_complex}).

The coding and discrimination capabilities of every individual neuron can be seen best in the subplots of Figs.~\ref{fig:LL_population_complex}A and \ref{fig:LL_population_complex}B. Neurons 1 and 2 both coded for the first four of the eight stimuli, but neuron 1 seemed to respond to one feature common in all of these stimuli whereas neuron 2 responded to each of these stimuli in a different way (Fig.~\ref{fig:LL_population_complex}A). This implies a sensitivity to four different features that were present in just one of these stimuli each. Because of this, only neuron 2 was able to distinguish between these four stimuli themselves, whereas both neurons were able to discriminate between any of the first four stimuli and any of the last four (compare $M_1$ and $M_2$ in Fig.~\ref{fig:LL_population_complex}B). Neurons 3 and 4  were each sensible to two of these four features (neuron 3 equally and neuron 4 differently) and neurons 5 and 6 to just one. So among the fist six neurons there was a hierarchy of specificity with neurons 1 and 2 the least and neurons 5 and 6 the most specific.
All these neurons also showed different levels of reliability, which caused that only three of them (neurons 1, 2 and 4) obtained maximum discrimination values for some stimulus pairs (Fig.~\ref{fig:LL_population_complex}, C and D) and thus contributed to the optimal LL population (Fig.~\ref{fig:LL_population_complex}E). An even higher level of redundancy could be observed for neurons 7 and 8, which both coded exactly for the same two stimuli, but here only neuron 8 was reliable enough to enter the optimal LL population. Neuron 9 did not respond to any of the stimuli, i.e. this neuron was either just noisy or sensitive only to one (or more) feature(s) that were not present in any of the stimuli.

In the simpler example of Fig.~\ref{fig:LL_population}, all off-diagonal elements of the population performance matrix $P$ were positive which means that all pairs of stimuli could successfully be discriminated. The optimized LL population consisted of three neurons, one vehicle type neuron and two color neurons, which came about because both color neurons happened to be most discriminative for some of the stimulus pairs. Here it would  take only two neurons to perfectly discriminate, as long as a color neuron was combined with a vehicle type neuron. This nicely illustrates the subtle distinction between coding and discrimination: with only the 'white' neuron 1 and the 'car' neuron 3 all stimuli could be discriminated even though none of these two neurons actually coded for stimulus 4, the 'red ship'. While this contained a case of discrimination without coding, exactly the opposite case occurred in Fig. \ref{fig:LL_population_complex}. There, for stimulus pair 7 and 8 we obtained a zero outside of the diagonal indicating that these two stimuli could not be discriminated by this set of neurons, although neuron 10 was responding to both of these stimuli. Therefore, this is a case of coding without discrimination.
In general, two stimuli cannot be discriminated whenever the population contains only neurons that are not sensitive to any of the features that distinguish them. Some of the neurons might still code the stimuli but just not the distinction. In Fig.~\ref{fig:LL_population}, a red car and a red ship could not be discriminated if only 'color' neurons were evaluated.

%
%

\section{\label{s:Discussion}Discussion}

\subsection{\label{ss:Discussion-Summary} Summary}

We evaluated methods for identifying the most discriminative subpopulation when looking at neuronal responses to repeated presentations of a set of stimuli. Central to our studies were two different assumptions about neuronal population coding. According to the {\em{summed population (SP) hypothesis}} all neurons in a population potentially contribute to the coding of an external stimulus, whereas under the {\em{labeled line (LL) hypothesis}} stimulus encoding is realized through individual neurons. In both cases our analysis relied on the computation of pairwise distance matrices and a basic discrimination performance that quantifies the degree to which identical (different) stimuli yield similar (dissimilar) spike train responses. However, since the two hypotheses include different assumptions about neuronal coding they required complementary algorithmic approaches.

For SP we compared three approaches that search the space of all possible subpopulations for the maximum discriminative performance over all stimulus pairs: (i) A {\em{brute force}} search evaluating all possible subpopulations; (ii) two variants of a steepest ascent or {\em{gradient algorithm}}; and (iii) {\em{simulated annealing}}. By definition, approach (i) provides the best results as evaluating all possible combinations guarantees that the global maximum is found. However, the number of possible subpopulations increases exponentially with the number of neurons (Eq.~\ref{eq:Total-number-of-combinations}), rendering a brute force approach feasible only for rather small populations. Gradient algorithms (ii), like the one used in \citet{Ince13}, overcome this limitations, as the number of subpopulations that has to be evaluated increases only quadratically with the number of neurons (Eq.~\ref{eq:Selected-number-of-combinations}). Unfortunately, while useful under very idealized conditions, in general these approaches have problems in finding the correct solution. In contrast, our simulated annealing approach (iii) proved much more reliable in finding the maximal discriminative power than both variants of the gradient approach, and this despite only a moderate increase in computational cost in the majority of cases studied here (see Fig. \ref{fig:Speed} for an example).

For LL we introduced an optimization algorithm which finds the most discriminative subpopulation by evaluating every neuron separately. First, for every individual stimulus pair the algorithm identifies the discriminative neurons and selects the best one. These best neurons are then combined to form the optimal LL-population. As shown in Fig.~\ref{fig:LL_population_complex}, the algorithm can handle quite involved coding scenarios, even though its computational complexity is much lower than in the SP case (because we only have to deal with one distance matrix per neuron). Moreover, we are guaranteed to find the best subpopulation since this time no search in a very high-dimensional subpopulation space is needed.

\subsection{\label{ss:Discussion-Comparisons} Comparisons}

To provide a more intuitive understanding of the relationship between the SP and the LL hypothesis, consider a single neuron that in itself is able to discriminate the whole set of stimuli perfectly. Such a neuron is where both hypothesis meet. It could be considered either as a perfect SP population of size one or as an LL neuron that yields a perfect discrimination matrix on its own. While for a few stimuli one neuron alone might indeed be able to do the job, the distinction of a large and complex set of stimuli typically works along several feature dimensions and a population of neurons is needed to work in unison and complement each other. Conversely, in order to obtain a robust and universal subpopulation in an experimental setting, it is essential to test the population on a stimulus set that is as diverse as possible. \highlight{In general, for both the SP and the LL case one should always keep in mind that the results obtained are a function of both the stimulus set presented and the neuronal subpopulation recorded.}

For such a complex stimulus set, SP and LL hypotheses can be seen as two distinct ways neurons as a population may collaborate to carry out the task of the perfect neuron. In the SP case they divide the spikes of each individual response among themselves, while in the LL case the responses to different stimuli are distributed among the neurons. It would also be possible to construct all kind of SP and LL mixtures by combining subdivision of spikes among neurons (SP coding blocks) with separation of stimulus sensitivity between different neurons (LL subpopulations) at any level. However, since this can get arbitrarily complex, a modification of the algorithms to address these mixed cases seems to be computationally out of reach.

There are other potential modifications. The current LL algorithm identifies for each stimulus pair all the neurons that are sensitive to discrimination and selects only the very best one. Overall this will result in a rather minimal LL-subpopulation. However, in real-life applications additional criteria such as redundancy and reliability might be very important. To create more stability it could be preferable to include more rather than less neurons \citep{Sanchez04}. This could be particularly useful whenever two neurons distinguish the same stimulus pair but use different features to do so. The inclusion of both of these neuron would guarantee access to complementary information that might be essential to discriminate more diverse stimulus sets. Then, both algorithms are designed to identify the most \textit{discriminative} subpopulation which is why we always look at pairs of two stimuli. If instead one were interested in finding the \textit{coding} subpopulation, the algorithms could be modified to evaluate responsiveness to individual stimuli.

\highlight{For the sake of simplicity we restricted our simulations to cases with uniform rates. However, recent studies on the observed variability in firing rates have emphasized the prevalence of log-normal distributions \citep{Buzsaki14} and the roles of 'soloists and choristers' \citep{Okun15}. This raises the question of the relative importance of coding by sparse-firing vs. higher-frequency neurons.  In the SP case varying firing rates may imply that the population spike train is divided among the population with a non-uniform probability distribution. However, this will only change the gradient towards the optimal solution and not the solution itself. Thus, our algorithm will not be affected. Likewise, in the LL case, the stimulus pairs are assessed individually and therefore the rate is less important. The neurons that display the most consistently distinct responses for different stimuli will be selected regardless of the actual rates.}
 
Our algorithms employ the SPIKE-distance as a fundamental measure for comparing spike trains. The only other approaches that use spike train distances to evaluate the coding properties of neuronal ensembles are population extensions for the Victor-Purpura distance \citep{Aronov03} and the van Rossum distance \citep{Houghton08}. These population measures interpolate between the two extreme cases of summed population and labeled line coding by means of a parameter that determines the importance of distinguishing spikes fired in different cells (minimal for SP, maximal for LL). While the applications to the pooled spike train of the full population (SP) or to all individual spike trains separately (LL) are straightforward (\citeauthor{Schneider10}, \citeyear{Schneider10} resp. \citeauthor{Mackevicius12}, \citeyear{Mackevicius12}), it is not obvious how to interpret intermediate cases. 
More importantly, these approaches never deal with subpopulations but always consider the population as a whole. This hampers comparison of results because these population extensions have not been designed to answer questions about the extent to which a part of the population contributes to stimulus discrimination and much less to identify the most discriminative subpopulation. In short, our algorithms and these population extensions address complimentary questions.

\subsection{\label{ss:Discussion-Outlook} Outlook}

So far we have applied both the SP and the LL algorithms to simulated datasets only. We either knew the ground truth a priori or we could apply the brute force approach to obtain the ground truth. After having passed this verification test, the next step will be to analyze experimental datasets in which a neuronal population is recorded upon repeated presentations of a set of stimuli. These can be data from animal models in both sensory and motor regions, but also recordings of intracranial neuronal spiking from patients undergoing seizure monitoring prior to epilepsy surgery \citep{Fried97}. 

\highlight{Once the algorithms are applied to experimental data their results may serve to address further fundamental questions about neuronal coding. One of them regards the size of the most discriminative subpopulation and how it compares to the size of the full population and to the size of the subpopulation that conveys the same information as the whole (see, e.g., the information theoretic analysis carried out in Ince et al., 2013). Moreover, it would be interesting to investigate the spatial location of the discriminative neurons and search for properties that distinguish them from other neurons. For example, one may evaluate their overall firing rates and their level of connectivity \citep{Buzsaki14} as well as their coupling to the overall firing of the population \citep{Okun15}.}

Another potential application for our algorithms are BCIs (see, e.g., \citeauthor{Lebedev06}, \citeyear{Lebedev06}; \citeauthor{Mak09}, \citeyear{Mak09}), in particular, the kind of BCI that works with multi-unit spike train read-out \citep{Homer13}. Current BCI systems are following the so-called mass-effect principle and rely on rather crude population averages like mean firing rates over many neurons \citep{Nicolelis09}. This could be improved significantly by increasing the signal-to-noise ratio (SNR) by selecting the most informative neurons \citep{Sanchez04} and making explicit use of the temporal structure of spike trains \citep{Sanchez08}. Algorithms for finding the most discriminative subpopulation based on the spike timing sensitive SPIKE-distance could lead to refined and more targeted estimates of ensemble activity. Comparing the SP and LL algorithms on the same dataset may provide further insights on how neural circuits encode. In fact, the success or failure of clinical BCI applications will depend on our efforts to understand population coding.

\highlight{To facilitate such efforts we made all the algorithms used here freely available on our download page\footnote[1]{http://www.fi.isc.cnr.it/users/thomas.kreuz/Source-Code/subpopulations.html} (Matlab command line).}

\section*{Appendix: The SPIKE-Distance} \label{App-s:SPIKE-Distance}

The spike train distance that we use to evaluate whether the responses elicited by different stimuli can be distinguished is the SPIKE-distance (\citeauthor{Kreuz11}, \citeyear{Kreuz11}; \citeauthor{Kreuz13}, \citeyear{Kreuz13}, see \citeauthor{Satuvuori17}, \citeyear{Satuvuori17} for a generalized version).
In contrast to the ISI-distance \citep{Kreuz07c, Kreuz09}, the SPIKE-distance is sensitive to both firing rate and spike timing. While spike-resolved distances like the Victor-Purpura distance or the van Rossum distance rely on a time-scale parameter, the time-resolved SPIKE-distance is parameter-free and time-scale independent. This allows for easy comparability of results obtained for vastly different firing rates, e.g. for pooled neuronal populations of different sizes \citep{Satuvuori18}. \highlight{But note also that for datasets with very low spike counts ($\leq 4$ spikes) other spike train distances might be preferable \citep{Satuvuori18}.}.

The SPIKE-distance $D$ measures the relative spike timing between spike trains normalized to local firing rates.
In order to assess the accuracy of spike events, each spike is assigned the distance to its nearest neighbor in the other spike train
\begin{equation} \label{eq:Closest spike}
	\Delta t_i^{(n)} = \min_j\left(\left|t_i^{(n)} - t_j^{(m)}\right|\right).
\end{equation}
These distances are interpolated between spikes using for all times $t$ the time differences
to the previous spike
\begin{equation} \label{eq:Weighting_prev}
	x_P^{(n)}(t) = t-t_i^{(n)}    \quad \textrm{for }\quad  t_i^{(n)} \leqslant t \leqslant t_{i+1}^{(n)},
\end{equation}
and to the following spike
\begin{equation} \label{eq:Weightings_foll}
	x_F^{(n)}(t) = t_{i+1}^{(n)}-t \quad \textrm{for }\quad  t_i^{(n)} \leqslant t \leqslant t_{i+1}^{(n)}.
\end{equation}
This defines a time-resolved dissimilarity profile from discrete values. The instantaneous weighted spike time difference for a spike train can be determined via the interpolation from one difference to the next
\begin{equation} \label{eq:weighted distance}
	S_{n} (t) = \frac{\Delta t_i^{(n)} (t)x_F^{(n)}(t) + \Delta t_{i+1}^{(n)}
		(t)x_P^{(n)}(t)}{x_{\mathrm{ISI}}^{(n)} (t)}\ ,\quad  t_i^{(n)} \leqslant t \leqslant
		t_{i+1}^{(n)}.
\end{equation}
The pairwise SPIKE-distance profile is subsequently obtained by averaging the weighted spike time differences, normalizing to the local firing rate average, and weighting each profile by the instantaneous firing rates of the two spike trains
\begin{equation} \label{eq:SPIKE profile}
	S_{m,n} (t) = \frac{S_n x_{\mathrm{ISI}}^m (t) +S_m x_{\mathrm{ISI}}^n (t) }
		{2\left\langle x_{\mathrm{ISI}}^{n,m} (t)\right\rangle^2 }.
\end{equation}
Averaging over all pairwise SPIKE-profiles results in the multivariate SPIKE-profile
\begin{equation} \label{eq:multivariate}
	S (t) = \tfrac{2}{N(N-1)}\sum_{m=1}^{N-1} \sum_{n=m+1}^N S_{m,n} (t).
\end{equation}
Finally, integration over time gives the distance value
\begin{equation} \label{eq:distance}
	D=\tfrac{1}{t_e-t_s} \int\limits_{t_s}^{t_e} S (t) dt .
\end{equation}
Here, $t_s$ and $t_e$ denote the start and end of the recording, respectively. The dissimilarity profile $S (t)$ and the SPIKE-distance $D$ are bounded to the interval $[0, 1]$. The distance value $D = 0$ is obtained for identical spike trains only.

Implementations of the SPIKE-distance are provided online in three separate freely available code packages called SPIKY\footnote[2]{http://www.fi.isc.cnr.it/users/thomas.kreuz/Source-Code/SPIKY.html} (Matlab graphical user interface, \citeauthor{Kreuz15}, \citeyear{Kreuz15}), PySpike\footnote[3]{http://mariomulansky.github.io/PySpike/} (Python library, \citeauthor{Mulansky16}, \citeyear{Mulansky16}) and cSPIKE\footnote[4]{http://www.fi.isc.cnr.it/users/thomas.kreuz/Source-Code/cSPIKE.html} (Matlab command line with MEX-files).

\vspace{1cm}

\section*{\label{s:Acknowledgement} \textbf{Acknowledgements}}
We thank Ralph G. Andrzejak, Nebojsa Bozanic, Irene Malvestio and Florian Mormann for many useful discussions.
This project has received funding from the European Union’s Horizon 2020 research and innovation programme under the Marie Sk\l{}odowska-Curie grant agreement No 642563, 'Complex Oscillatory Systems: Modeling and Analysis' (COSMOS).



\bibliographystyle{elsarticle-harv}\biboptions{authoryear}

\end{document}